\documentclass[a4paper,11pt]{article}
\pdfoutput=1

\usepackage{amsbsy}
\usepackage{amsfonts}
\usepackage{amsmath}
\usepackage{amssymb}
\usepackage{bm}
\usepackage{enumerate}
\usepackage[T1]{fontenc}
\usepackage{graphicx}
\usepackage[utf8]{inputenc}
\usepackage{jcappub}
\usepackage{mathtools}
\usepackage{pstricks}
\usepackage{setspace}
\usepackage[normalem]{ulem}
\usepackage{xcolor}

\DeclareMathOperator\erf{erf}
\DeclareMathOperator{\Tr}{Tr}

\newcommand{\SNR}{{\rm SNR}}

\newcommand{\bn}{{\bm n}}
\newcommand{\de}{{\rm d}}
\newcommand{\euc}{\textit{Euclid}}
\newcommand{\tj}[6]{ \begin{pmatrix}
  #1 & #2 & #3 \\
  #4 & #5 & #6
\end{pmatrix}}
\newcommand{\Gj}[6]{ \begin{Bmatrix}
  #1 & #2 & #3 \\
  #4 & #5 & #6
\end{Bmatrix}}
\title{Speeding up the detectability of the harmonic-space galaxy bispectrum}

\author[a]{Francesco Montanari}
\author[b,c]{and Stefano Camera}

\affiliation[a]{Instituto de F\'isica Te\'orica IFT-UAM/CSIC, Universidad
  Aut\'onoma de Madrid,\\Cantoblanco 28049 Madrid, Spain}
\affiliation[b]{Dipartimento di Fisica, Universit\`a degli Studi di Torino,\\Via P.\ Giuria 1, 10135 Torino, Italy}
\affiliation[c]{INFN -- Istituto Nazionale di Fisica Nucleare, Sezione di Torino,\\Via P.\ Giuria 1, 10135 Torino, Italy}
\emailAdd{francesco.montanari@uam.es}
\emailAdd{stefano.camera@unito.it}

\abstract{We present a method that allows us for the first time to estimate
  the signal-to-noise ratio (SNR) of the harmonic-space galaxy bispectrum
  induced by gravity, a complementary probe to already well established
  Fourier-space clustering analyses. We show how to do it considering only
  $\sim1000$ triangle configurations in multipole space, corresponding to a
  computational speedup of a factor $\mathcal{O}(10^2)-\mathcal{O}(10^3)$,
  depending on the redshift bin, when including mildly non-linear scales.
  Assuming observational specifications consistent with forthcoming
  spectroscopic and photometric galaxy surveys like the \euc\ satellite and
  the Square Kilometre Array (phase 1), we show: that given a single redshift
  bin, spectroscopic surveys outperform photometric surveys; and that---due to
  shot-noise and redshift bin width balance---bins at redshifts $z\sim1$ bring
  higher cumulative SNR than bins at lower redshifts $z \sim 0.5$. Our results
  for the largest cumulative $\SNR \sim 15$ suggest that the harmonic-space
  bispectrum is detectable within narrow ($\Delta z \sim 0.01$) spectroscopic
  redshift bins even when including only mildly non-linear scales. Tomographic
  reconstructions and inclusion of highly non-linear scales will further boost
  detectability with upcoming galaxy surveys. In addition, we discuss how,
  using the Karhunen-Lo\`eve transform, a detection analysis only requires a
  $1 \times 1$ covariance matrix for a single redshift bin.}

\begin{document}
\maketitle
\flushbottom

\section{Introduction}

The clustering of galaxies is one of the most important cosmological
probes. Hitherto, it has been explored mostly through its two-point
statistics, like the galaxy correlation function or the galaxy power
spectrum. Both methods have provided excellent constraints on
cosmological parameters
\cite[e.g.][]{Abbott:2017wau,Camacho:2018mel,Alam:2020sor}, soon to be
boosted by upcoming surveys that will cover unprecedented volumes and
source number densities. In particular, it is worth mentioning: the
European Space Agency's flagship, the \euc\ satellite
\citep{Laureijs:2011gra,Amendola:2012ys,Amendola:2016saw,Blanchard:2019oqi};
the Rubin Observatory (previously known as Large Synoptic Survey
Telescope, LSST) \citep{Ivezic:2008fe,Abate:2012za}; the Dark Energy
Spectroscopic Instrument
\citep[DESI;][]{Flaugher:2014lfa,Aghamousa:2016zmz,Levi:2019ggs}; and
the Square Kilometre Array
\citep[SKA;][]{Maartens:2015mra,Bull:2015esa,Raccanelli:2015hsa,Abdalla:2015zra,Santos:2015bsa,Bacon:2018dui}.

On the other hand, due to both a more complex modelling and the
limitations of previously available data sets, as well as to high
computational requirements, higher-order summary statistics such the
bispectrum have played a minor role up to now. However, it is well
known that the bispectrum (or its Fourier transform, the 3-point
correlation functions) represents a unique window to the primordial
Universe \cite{Celoria:2018euj} and a complementary probe of the
large-scale structure (LSS)
\cite{2015MNRAS.451..539G,Gil-Marin:2016wya,2017MNRAS.469.1738S,2017MNRAS.468.1070S,2018MNRAS.478.4500P,Yankelevich:2018uaz,Rizzato:2018whp}.

Often, most studies of galaxy clustering poly-spectra (i.e.\ power spectrum and beyond) are carried out in Fourier
space \cite{Bernardeau:2001qr}. However, Fourier-space analyses require a fiducial cosmological model to convert
observed angles and redshifts into physical distances and, thus, into Fourier wave-numbers. Instead, the dependence on multipoles
and redshifts in harmonic space allows us to perform analyses independent
of the dynamics of a specific cosmological model. While the conversion of measurements on the celestial sphere into Fourier modes is typically performed iteratively, and it is controlled by
consistency tests, the harmonic- and Fourier-space estimators are
complementary probes and tensions in cosmological parameter
constraints obtained with the two methods may be relevant to explain
e.g.\ recent literature results pointing to tensions in determinations
of the Universe expansion rate \cite{Camarena:2019rmj}. Furthemore,
harmonic-space statistics has a different dependence on some
systematic errors than what happens in Fourier space. Hence, here we consider
harmonic space clustering analyses not only in view of upcoming
photometric galaxy catalogues (for which the relatively poor redshift
determination hinders Fourier-space analyses), but also in view of
spectroscopic catalogues (typically analysed in Fourier space).

Source number counts have been computed in several perturbation
schemes and at different orders in the past (see
Ref.~\cite{Bernardeau:2001qr} for a review in the context of standard
perturbation theory). In this work, we rely on the formalism developed
in Refs.~\cite{DiDio:2014lka,DiDio:2015bua,DiDio:2018unb} for the
tree-level harmonic bispectrum, valid for arbitrary non-interacting
dark energy models and modified gravity models provided that photons
and dark matter particles move along geodesics. Inclusion of radial
selection functions has proven to be computationally challenging even
for a simple estimate of the cumulative signal-to-noise ratio (SNR)
due to the large number of modes. Here, we provide methodologies to
estimate the cumulative SNR using only $\sim 10^3$ multipole
configurations, compared to the total
$\mathcal{O}(10^5)-\mathcal{O}(10^6)$ available within single redshift
bins consistent with upcoming photometric measurements like those
performed by \euc\ or the Rubin Observatory, or spectroscopic
observations like for DESI, SKA surveys, or, again, \euc.

Given a methodology for the computation of the theoretical bispectrum,
we further discuss a possible strategy for efficient data fit in a
detection analysis. Optimal bispectrum estimators and related
efficient computational strategies can be adapted from works developed
for Cosmic Microwave Background studies
\cite[e.g.][]{Coulton:2017crj,Babich:2004gb,Creminelli:2005hu}.
However, here we are interested in reaching a drastic dimensionality
reduction. Indeed, the covariance matrices needed for the fit can be
either computed theoretically (as done here) or estimated from
simulations. The latter option is extremely computationally expensive,
as the number of simulations needs to be larger than the number of
elements of the data vector, which itself is
$\mathcal{O}(10^5)-\mathcal{O}(10^6)$ \cite{Gualdi:2017iey}.
Oppositely, in our approach the covariance computational runtime is
negligible compared to the bispectrum one, but data fitting will still
require comparison with simulations to asses the validity of our
assumptions. For instance, simulations are necessary to validate the
smallest scale included in the analysis. Also, finite volume effects
will introduce multipole correlations---here neglected---that in the
case of the power spectrum may be mitigated, e.g. via
multipole binning validated comparing analytical estimates to
simulations \cite{Cabre:2007rv}. Therefore, we show how to apply the
Karhunen-Lo\`eve transform (KLT) \cite{Tegmark:1996bz} to the
tree-level spherical harmonic bispectrum. The KLT has been used for
the Fourier-space bispectrum in \cite{Gualdi:2017iey} to compress
information in wave-numbers, and for the harmonic-space power spectrum
in \cite{Alonso:2017hhj} to compress radial information in a
tomographic analysis involving correlations between several redshift
bins. Here, we are rather interested in compressing information in
multipoles, since the large number of physical non-vanishing
triangular configuration makes it prohibitive to simulate covariance
matrices already for a single redshift bin.

In \autoref{sec:bisp}, we review fundamental results for the
tree-level harmonic-space bispectrum and its variance. In
\autoref{sec:geom}, we study geometric properties of the bispectrum
SNR in multipole space. Forecasts specifications are given
in \autoref{sec:specs}, while the forecast methodology and results
are presented in \autoref{sec:forecast}, and \autoref{sec:kl}
discusses how to reach efficient parameter constraints via the
Karhunen-Lo\`eve transform. We conclude in
\autoref{sec:conclusions}. In \autoref{sec:geomfact}, we list
geometrical factors relevant for the bispectrum computation. In
\autoref{sec:interp}, we study an alternative forecast methodology as
a consistency check for the main analysis. In
\autoref{sec:numeric}, we give details about the numerical
computation of the bispectrum SNR.

Our fiducial cosmology throughout this paper is a flat $\Lambda$CDM
model with Hubble parameter, dark matter and baryon density
parameters, amplitude, tilt and pivot of the primordial power spectrum
given by: $\{h=0.67, \Omega_{\rm cdm}=0.27, \Omega_{\rm b}=0.05,
A_{\rm s}=2.3 \times 10^{-9}, n_{\rm s}=0.962, k_*=0.05/{\rm Mpc} \}$.

\section{Tree-level bispectrum and its variance}
\label{sec:bisp}
We consider the tree-level bispectrum formalism developed in
\cite{DiDio:2014lka,DiDio:2015bua,DiDio:2018unb}. We verified that,
given the wide redshift bins considered here, redshift-space
distortions and other local terms discussed in \cite{DiDio:2018unb}
are safely negligible. Given that we do not consider correlations
among different redshift bins, also integrated terms (e.g.\ lensing)
are negligible for our purposes. Such terms could be relevant for the
auto-correlation of a single bin only if this extends over a much
larger range $\Delta z \sim \mathcal{O}(1)$ than those of our interest
$\Delta z \sim \mathcal{O}(0.1)$ \cite{DiDio:2015bua}. Hence, for the
purposes of our forecasts the bispectrum induced by gravitational
non-linearities is well-approximated by the dominant density
contribution.

We assume that source density perturbations are related to matter
density perturbations $\delta$ via a local bias model, neglecting
stochastic bias terms
\begin{equation}
  \label{eq:11}
  \delta_g = b_1\delta+{1\over2}b_2\, \delta^2 + b_{s^2}\, s^2 \,.
\end{equation}
We assume the bias coefficients $b_1$, $b_2$, $b_{s^2}$ to be
scale-independent. The bias coefficient $b_{s^2}$ is related to the tidal
field $s_{ij}$ \cite{Desjacques:2016bnm} and we defined
$s^2=s_{ij} s^{ij}$. We expand perturbations up to second-order terms
$\delta = \delta^{(1)}+\delta^{(2)}$.

The bispectrum of density fluctuations is defined as
\begin{multline}
  \label{eq:dens_corr}
  B^{\delta^{(2)}}(\bn_1,\bn_2,\bn_3,z_1,z_2,z_3) =\\
  \Bigg\langle \left\{ b_1(z_1) \delta^{(2)}( \bn_1, z_1 ) +
     \frac{b_2(z_1)}{2} \left[\delta^{(1)}( \bn_1, z_1 )\right]^2 +
     b_{s^2}(z_1) s^2\left( \bn_1, z_1 \right) \right\}\\
     \times \left[ b_1(z_2) \delta^{(1)} ( \bn_2, z_2 ) \right]
     \left[ b_1(z_3) \delta^{(1)} ( \bn_3, z_3 ) \right]
     + \circlearrowleft \Bigg\rangle \;,
\end{multline}
where $\circlearrowleft$ denotes two additional cyclic permutations
over the arguments $(\bn_i, z_i)$. These parameters represent the
direction of observation $-\bn_i$ and the redshift $z_i$ of a given
source. The bispectrum can be expanded in spherical harmonics
\begin{equation}
  \label{eq:10}
  B(\bn_1,\bn_2,\bn_3,z_1,z_2,z_3) =
  \sum_{\substack{\ell_1, \ell_2, \ell_3\\ m_1, m_2, m_3}}
  B^{m_1m_2m_3}_{\ell_1\ell_2\ell_3}(z_1,z_2,z_3)
  Y_{\ell_1m_1}(\bn_1) Y_{\ell_2m_2}(\bn_2) Y_{\ell_3m_3}(\bn_3) \;,
\end{equation}
and, using statistical isotropy, the physical information can be
further factorised in terms of the reduced bispectrum defined by
\begin{equation}
  \label{eq:6}
  B^{m_1m_2m_3}_{\ell_1\ell_2\ell_3}(z_1,z_2,z_3) =
  \mathcal{G}^{m_1m_2m_3}_{\ell_1\ell_2\ell_3}
  b_{\ell_1\ell_2\ell_3}(z_1, z_2, z_3) \;.
\end{equation}
In \autoref{sec:geomfact} we define the Gaunt integral
$\mathcal{G}^{m_1m_2m_3}_{\ell_1\ell_2\ell_3}$, which is zero unless
$m_1+m_2+m_3 = 0$ and the following multipole conditions hold:
\begin{align}
  \label{eq:triangle-1}
  &&|\ell_2-\ell_3| \leq \ell_1 \leq \ell_2 + \ell_3
     \quad \mbox{(triangle inequality)} \;,\\
  \label{eq:triangle-2}
  &&\ell_1 + \ell_2 + \ell_3 = {\rm even} \;.
\end{align}
The triangle inequality must be satisfied for all indices
permutations.

Using standard cosmological perturbation theory at tree-level, the
reduced bispectrum can be written in terms of generalised harmonic
power spectra
\begin{equation}
  \label{eq:c_ll}
  \prescript{n}{}{C}_{\ell\ell^\prime}(z_1,z_2)= i^{\ell-\ell^\prime} 4\pi
  \int \de\ln k\, k^n \mathcal{P}_R \left( k \right)
  \Delta_\ell \left( k , r_1 \right) \Delta_{\ell^\prime} \left( k , r_2 \right)\,.
\end{equation}
Here $\mathcal{P}_R \left( k \right)$ is the dimensionless power
spectrum of primordial curvature perturbations, and we defined
$\Delta_\ell \left( k , r \right) = T_\delta \left( k , r \right)
j_\ell \left( k r \right)$, where $T_\delta \left( k , r \right)$ is
the linear transfer function of density perturbations
\cite{DiDio:2013bqa}, $j_\ell(x)$ is the spherical Bessel function and
$r(z)$ is the radial comoving distance to redshift $z$. The reduced
bispectrum reads
\begin{multline}
 b^{\delta^{(2)}}_{\ell_1 \ell_2 \ell_3}(z_1,z_2,z_3)
=\left[ b_1(z_1) + \frac{21}{34} b_2 (z_1) \right] b^{\delta0 }_{\ell_1 \ell_2 \ell_3 } (z_1,z_2,z_3)
 + b_1 (z_1)  b^{\delta1}_{\ell_1 \ell_2 \ell_3 }(z_1,z_2,z_3)
\\+
 \left[ b_1(z_1) + \frac{7}{2} b_{s^2} (z_1)
 \right] b^{\delta2}_{\ell_1 \ell_2 \ell_3 } \left( z_1 , z_2 ,
   z_3 \right)  {+\circlearrowleft}\;,
\end{multline}
where we further defined the following contributions
\begin{itemize}
\item Monopole:
  \begin{equation}
    b^{ \delta0 }_{\ell_1 \ell_2 \ell_3 } (z_1,z_2,z_3)
    =
    \frac{34}{21}
    C_{\ell_1}(z_1,z_2) C_{\ell_2}(z_1,z_3)
    \,.
  \end{equation}

\item Dipole (the geometrical factors $g_{\ell_1\ell_2\ell_3}$ and
  $Q_{\ell\ \ell^\prime\ell^{\prime\prime}}^{\ell_1\ell_2\ell_3}$ are defined in
  \autoref{sec:geomfact}):
  \begin{equation}
    \label{eq:b_dens_dip}
    \begin{split}
    b^{ \delta1 }_{\ell_1 \ell_2 \ell_3 } (z_1,z_2,z_3) =
       \frac{ \left(g_{\ell_1\ell_2\ell_3}\right)^{-1}}{16\pi^2}
        \sum_{\ell^\prime\ell^{\prime\prime}}
      &  (2\ell^\prime+1)(2\ell^{\prime\prime}+1) Q_{1\ \ell^\prime
        \ell^{\prime\prime}}^{\ell_1 \ell_2 \ell_3}
      \\
    & \times \left[
       \prescript{1}{}{C}_{\ell^{\prime\prime} \ell_2}(z_1,z_2)
       \prescript{-1}{}{C}_{\ell^\prime \ell_3}(z_1,z_3)
       \right.
      \\
    & \qquad
       \left.+
       \prescript{-1}{}{C}_{\ell^{\prime\prime} \ell_2}(z_1,z_2)
       \prescript{1}{}{C}_{\ell^\prime \ell_3}(z_1,z_3)
       \right] \;.
    \end{split}
  \end{equation}
  $Q_{1\ \ell^\prime\ell^{\prime\prime}}^{\ell_1\ell_2\ell_3}$ is zero
  unless $\ell^\prime=\ell_2 \pm 1$ and $\ell^{\prime\prime}=\ell_1
  \pm 1$, hence the imaginary unit factors associated to generalized
  spectra lead to real results $i^{\ell^\prime+\ell^{\prime\prime}}
  (-i)^{\ell_1+\ell_2} = \pm 1$.
\item Quadrupole:
  \begin{equation}
    \label{eq:b_dens_quad}
    \begin{split}
    b^{\delta2 }_{\ell_1 \ell_2 \ell_3 } (z_1,z_2,z_3) =
    \frac{\left(g_{\ell_1\ell_2\ell_3}\right)^{-1}}{42\pi^2}
    \sum_{\ell^\prime\ell^{\prime\prime}}
    & (2\ell^\prime+1)(2\ell^{\prime\prime}+1) Q_{2\ \ell^\prime
      \ell^{\prime\prime}}^{\ell_1 \ell_2 \ell_3}
    \\
    & \times {C}_{\ell^{\prime\prime} \ell_2}(z_1,z_2) \
    {C}_{\ell^\prime \ell_3}(z_1,z_3) \;.
  \end{split}
  \end{equation}
  $Q_{2\ \ell^\prime\ell^{\prime\prime}}^{\ell_1\ell_2\ell_3}$ is zero unless
  $\ell^\prime=\ell_2\pm 2, \ell_2$ and $\ell^{\prime\prime}=\ell_1\pm 2, \ell_1$, hence
  $i^{\ell^\prime+\ell^{\prime\prime}} (-i)^{\ell_1+\ell_2} = \pm 1$.
\end{itemize}

The angle-averaged bispectrum (see \autoref{eq:aa-bisp}) covariance
for an arbitrary redshift-dependent angular bispectrum was computed in
\cite{DiDio:2018unb} in the Gaussian approximation
$B_{\ell_1\ell_2\ell_3}(z_1, z_2, z_3)\approx 0$. In this case the
covariance is diagonal, and the variance for the
$\ell_1+\ell_3+\ell_3=$ even case of our interest is given by
\begin{align} \label{eq:B_covarince_sumleven}
  &&\sigma^2_{B_{\ell_1\ell_2\ell_3}}(z_1, z_2, z_3) =
     C_{\ell_1}^{11}   C_{\ell_2}^{22}
      C_{\ell_3}^{33}
     + \left[
     C_{\ell_1}^{12}   C_{\ell_2}^{23}
      C_{\ell_3}^{31}
     +
     C_{\ell_1}^{13}   C_{\ell_2}^{21}
      C_{\ell_3}^{32}  \right]
     \delta_{\ell_1\ell_2} \delta_{\ell_2\ell_3}
     \nonumber \\
  && \quad +
     C_{\ell_1}^{11}   C_{\ell_2}^{23}
      C_{\ell_3}^{32}
     \delta_{\ell_2\ell_3}
     +
     C_{\ell_1}^{12}   C_{\ell_2}^{21}
      C_{\ell_3}^{33}
     \delta_{\ell_1\ell_2} +
     C_{\ell_1}^{13}   C_{\ell_2}^{22}
      C_{\ell_3}^{31}
     \delta_{\ell_1\ell_3} \;.
\end{align}
$\delta_{\ell_i\ell_j}$ is the Kronecker delta and we used the compact
notation $C_{\ell}^{ij} \equiv C_{\ell}(z_i, z_j) + \epsilon \delta_{ij}$,
where we introduced a Poisson shot-noise contribution
$\epsilon$.\footnote{See, e.g., \cite{Scoccimarro:2003wn} to compare this
  expression with the respective result in Fourier space.}

The observable bispectrum and its variance include integration over
radial selection functions $\phi_i(z)$:
\begin{equation}
  \label{eq:14}
  B_{\ell_1\ell_2\ell_3}^{ijk} = \int\de z_1\, \phi_i(z_1) \int\de z_2\, \phi_j(z_2) \int\de z_3 \,\phi_k(z_3) B_{\ell_1\ell_2\ell_3} (z_1, z_2, z_3) \;,
\end{equation}
\begin{equation}
  \label{eq:17}
  \sigma^2_{B_{\ell_1\ell_2\ell_3}^{ijk}} = \int\de z_1\, \phi_i(z_1) \int\de z_2\, \phi_j(z_2) \int\de z_3\, \phi_k(z_3) \sigma^2_{B_{\ell_1\ell_2\ell_3}}(z_1, z_2, z_3) \;.
\end{equation}

\section{Bispectrum geometry in multipole space}
\label{sec:geom}

To gain insights about geometrical properties of the bispectrum in
multipole space, in this section we neglect integration over radial
selection functions so that we can compute all the triangle
configurations satisfying \autoref{eq:triangle-1},
\autoref{eq:triangle-2}. We use the convention $3 \leq \ell \leq
\ell^\prime \leq \ell^{\prime\prime} \leq \ell_{\rm
  max}=200$.\footnote{The minimum multipole $\ell_{\rm min}=3$ is set
  by the fact that lower bispectrum multipoles depend on non-linear
  terms at the observer \cite{DiDio:2014lka}.} We consider the equal
redshifts case $z \equiv z_1=z_2=z_3=0.49$. These specific values of
$\ell_{\rm max}$ and $z$ correspond to the maximum multipole and the
mean redshift of our forecast lower photometric redshift bin (see
\autoref{sec:specs}, \autoref{sec:forecast})---we verified that the
picture is qualitatively the same in the range of our interest $z
\lesssim 1$, $\ell_{\rm max} \lesssim 300$. For this configuration we
can neglect shot-noise values of the same order of magnitude as those
used for our forecasts.

\begin{figure}[t]
  \centering
  \includegraphics[scale=0.47]{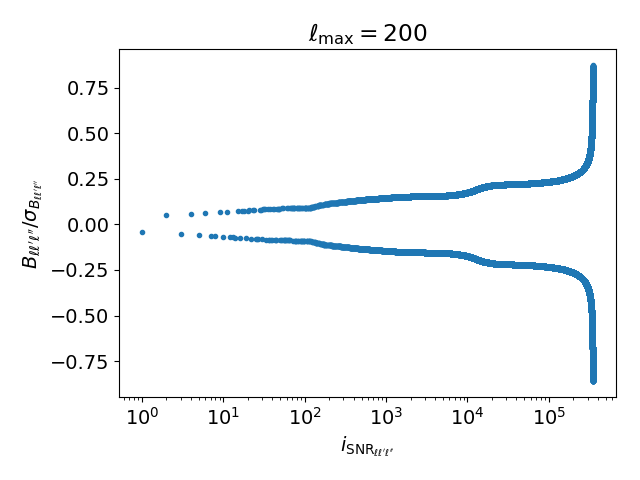}
  \includegraphics[scale=0.47]{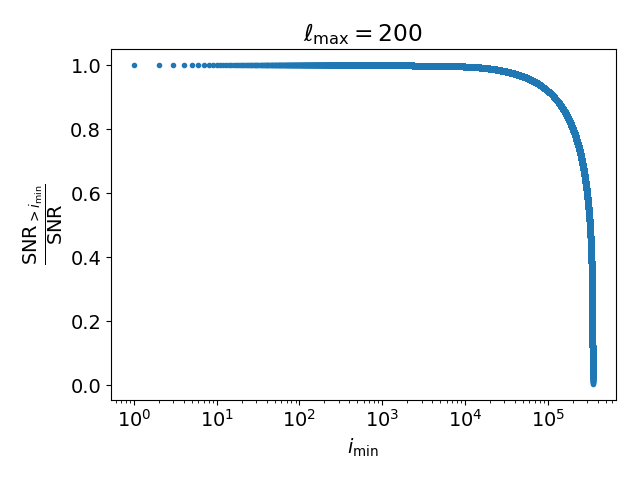}
  \caption{\emph{Left panel:} bispectrum over cosmic variance as
    function of the $i^{\rm th}$ multipole triangle. Triangles are
    ordered such that $\SNR_{\ell\ell^\prime\ell^{\prime\prime}}$ is
    monotonically sorted. \emph{Right panel:} Fraction of cumulative
    SNR obtained excluding the first $i_{\rm min}$ triangles.}
  \label{fig:bisp_cv}
\end{figure}

In \autoref{fig:bisp_cv} we show the bispectrum as a function of a
given multipole triangle over the respective cosmic variance,
$B_{\ell\ell^\prime\ell^{\prime\prime}} /
\sigma_{B_{\ell\ell^\prime\ell^{\prime\prime}}}$. The index
$i_{\SNR_{\ell\ell^\prime\ell^{\prime\prime}}}$ on the abscissa
identifies the triangles ordered to sort
$\SNR_{\ell\ell^\prime\ell^{\prime\prime}} =
|B_{\ell\ell^\prime\ell^{\prime\prime}}| /
\sigma_{B_{\ell\ell^\prime\ell^{\prime\prime}}}$. Let us note that the
symmetry around the abscissa would already allow us to estimate the
cumulative SNR\footnote{We take the sum only over $3 \leq \ell \leq
  \ell^\prime \leq \ell^{\prime\prime}$ rather than over $3 \leq \ell,
  \ell^\prime, \ell^{\prime\prime}$ because the bispectrum is
  invariant under permutations of multipole indices.}
\begin{equation}
  \label{eq:csn}
  \SNR\left(\le \ell_{\rm max}\right) = \sqrt{
    \sum_{3 \leq \ell \leq \ell^\prime \leq \ell^{\prime\prime} \leq \ell_{\rm max}}
    \frac{B^2_{\ell\ell^\prime\ell^{\prime\prime}}}{\sigma^2_{B_{\ell\ell^\prime\ell^{\prime\prime}}}}}
\end{equation}
considering only about half of the triangles, i.e.\ only those
triangles that lead to a positive bispectrum
$(B^+)_{\ell\ell^\prime\ell^{\prime\prime}}$ as $\SNR \approx \sqrt{2
  \sum (B^+)^2_{\ell\ell^\prime\ell^{\prime\prime}} /
  \sigma^2_{(B^+)_{\ell\ell^\prime\ell^{\prime\prime}}}}$ (similarly,
one could consider only negative bispectra) recovering the correct
value up to errors $\lesssim 0.1\%$. \autoref{fig:bisp_cv} also shows
the cumulative $\SNR_{>i_{\rm min}}$ obtained excluding the first
$i_{\rm min}$ triangles, relative to the total one. The first
triangles $i_{\rm min} \lesssim 10^3$ do not contribute significantly
and could be excluded from the SNR computation. However, in
\autoref{sec:forecast} we will use a more efficient approximation of
the cumulative SNR.

\begin{figure}[t]
  \centering
  \includegraphics[scale=0.35]{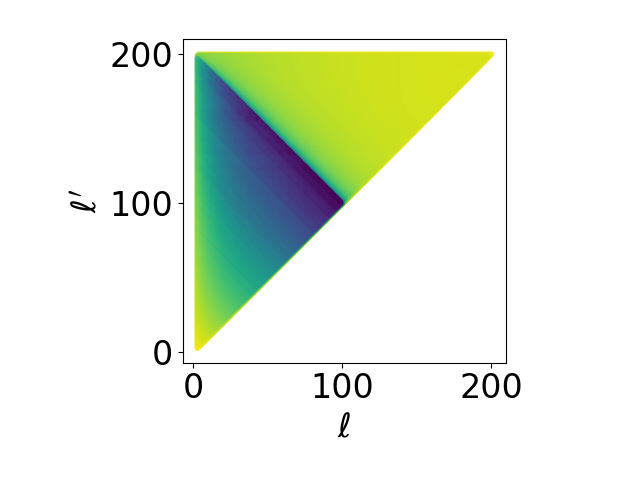}
  \hspace{-1.3cm}
  \includegraphics[scale=0.35]{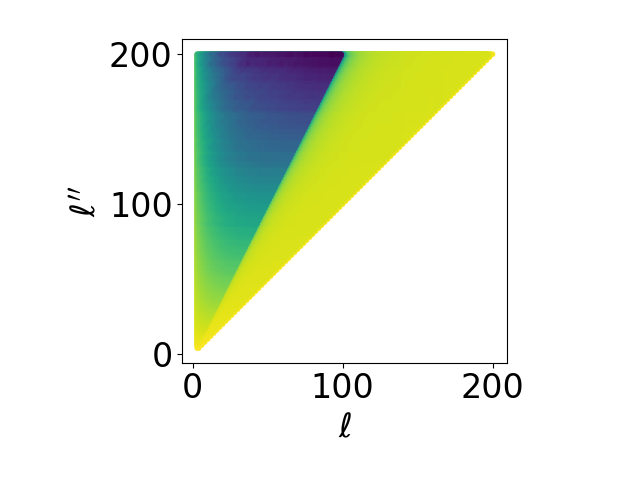}
  \hspace{-1cm}
  \includegraphics[scale=0.35]{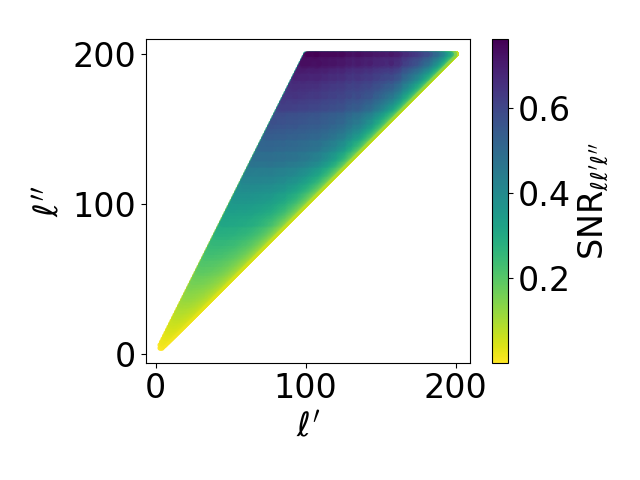}
  \caption{SNR as a function of multipoles. As many points overlap in
    these two-dimensional projections, we choose to show those with
    largest $\SNR_{\ell\ell^\prime\ell^{\prime\prime}}$ for each
    coordinate combination.}
  \label{fig:triangles}
\end{figure}

\begin{figure}[t]
  \centering
  \includegraphics[scale=0.3]{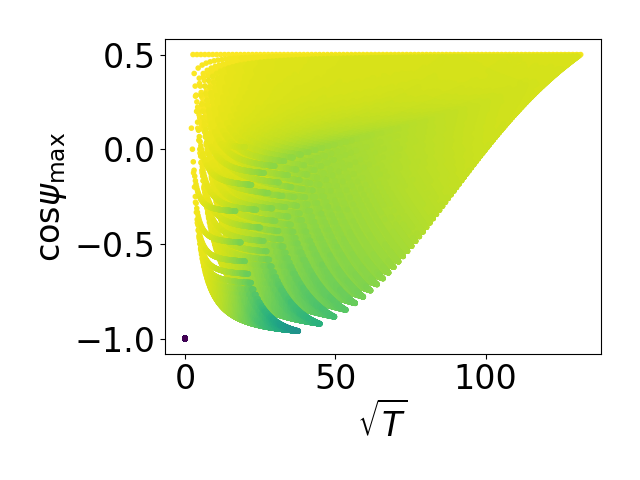}
  % \hspace{-1cm}
  \includegraphics[scale=0.3]{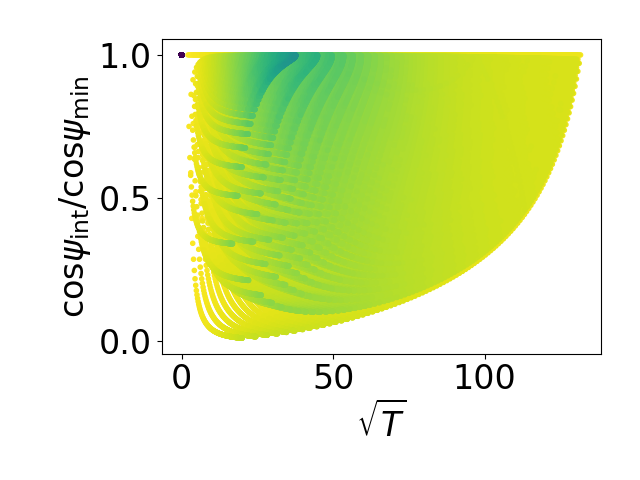}
  % \hspace{-1cm}
  \includegraphics[scale=0.3]{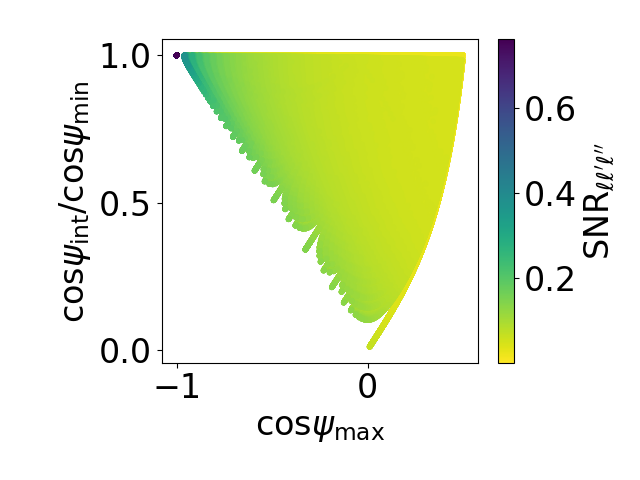}
  \caption{SNR as a function of the square root of the triangle's area
    $\sqrt{T}$, of the cosine of the largest internal angle $\cos
    \psi_{\rm max}$ and of the ratio between the cosines of the
    intermediate and smallest angles $\cos \psi_{\rm int} / \cos
    \psi_{\rm min}$.}
  \label{fig:geomax}
\end{figure}

\autoref{fig:triangles} shows the SNR per triangle,
$\SNR_{\ell\ell^\prime\ell^{\prime\prime}}$, as a function of
multipoles. The largest $\SNR_{\ell\ell^\prime\ell^{\prime\prime}}$
correspond to $\ell^\prime \approx \ell_{\rm max}-\ell$, peaking at
$\ell=\ell^\prime$, and to $\ell^{\prime\prime} \approx 2\ell$ for the
largest $\ell^{\prime\prime}$. Hence, the largest
$\SNR_{\ell\ell^\prime\ell^{\prime\prime}}$ corresponds to the folded
configuration $\ell\approx\ell^\prime\approx\ell^{\prime\prime}/2$.
Equilateral configurations $\ell=\ell^\prime=\ell^{\prime\prime}$
correspond to the minimum
$\SNR_{\ell\ell^\prime\ell^{\prime\prime}}$.\footnote{The apparent
  sharp transitions from large to small
  $\SNR_{\ell\ell^\prime\ell^{\prime\prime}}$ values at $\ell^\prime
  \approx \ell_{\rm max}-\ell$ in the $\ell-\ell^\prime$ plane, and at
  $\ell^{\prime\prime} \approx 2\ell$ in the
  $\ell-\ell^{\prime\prime}$ are misleading as overlapping points with
  smaller $\SNR_{\ell\ell^\prime\ell^{\prime\prime}}$ are not visible
  in these projections.}

As an alternative picture, rather than studying the dependence of
$\SNR_{\ell\ell^\prime\ell^{\prime\prime}}$ on the triangle side lengths, we consider
the following coordinates \cite{Gualdi:2019sfc}:
\begin{itemize}
  \label{test}
\item $\sqrt{T}$, the square root of the triangle's area.
\item $\cos \psi_{\rm max}$, the cosine of the largest internal angle.
\item $\cos \psi_{\rm int} / \cos \psi_{\rm min}$, the ratio between
  the cosines of the intermediate and smallest angles.
\end{itemize}
In \autoref{fig:geomax} several points with large
$\SNR_{\ell\ell^\prime\ell^{\prime\prime}}$ overlap at $\sqrt{T}=0$,
which, given our convention $\ell \leq \ell^\prime \leq
\ell^{\prime\prime}$, correspond to
$\ell+\ell^\prime=\ell^{\prime\prime}/2$; the further conditions $\cos
\psi_{\rm max}=-1$ and $\cos \psi_{\rm int} / \cos \psi_{\rm min}=1$
for the largest $\SNR_{\ell\ell^\prime\ell^{\prime\prime}}$ lead to
folded triangles $\ell \approx \ell^\prime \approx
\ell^{\prime\prime}/2$, as expected. These coordinates make it more
clear that equilateral triangles give the smallest
$\SNR_{\ell\ell^\prime\ell^{\prime\prime}}$, as this corresponds to
$\cos\psi_{\rm max}\approx1/2$ (i.e., $\psi_{\rm max}\approx60^\circ$)
along all $\sqrt{T}\neq0$ values, jointly with $\cos\psi_{\rm
  int}/\cos\psi_{\rm min}\approx1$.

\section{Forecast specifications}
\label{sec:specs}

In this section we outline observable specifications consistent with
upcoming galaxy surveys.

\subsection{Photometric survey}

We consider a photometric \euc-like survey \cite{Blanchard:2019oqi}.
Radial selection functions can be written as $\phi_i =W_i\, \de N/\de
z/\de\Omega$ \cite[e.g.][]{2012MNRAS.427.1891A}, where the galaxy
density per redshift and solid angle is
\begin{equation}
  \label{eq:18}
  \frac{\de N}{\de z\,\de\Omega}(z) = \left(\frac{z}{z_0}\right)^2
  \exp\left[-\left(\frac{z}{z_0}\right)^{3/2}\right] \;,
\end{equation}
with $z_0=z_{\rm m}/\sqrt{2}$ given the mean redshift $z_{\rm m}=0.9$, and
\begin{equation}
  \label{eq:8}
  W_i(z) = \int\de z_{\rm p}\,P(z_{\rm p}|z) \widetilde{W}_i(z_{\rm p}) \;.
\end{equation}
We assume a tophat selection $\widetilde{W}_i(z_{\rm p})$ in photometric redshift space
and we take a simple Gaussian form with standard deviation
$\sigma_{z, i}=0.05(1+\bar{z}_i)$ ($\bar{z}_i$ being the mean
photometric redshift within the $i$th bin) for the probability
$P(z_{\rm p}|z)$ that a galaxy with redshift $z$ has measured redshift
$z_{\rm p}$. Then the radial selection function is written in terms of the
error function as
\begin{equation}
  \label{eq:9}
  \phi_i(z) \propto \frac{\de N}{\de z\,\de\Omega} \left(\erf\left[\frac{z_i^+ -
        z}{\sqrt{2} \sigma_z}\right] - \erf\left[\frac{z_i^- -
        z}{\sqrt{2} \sigma_z}\right] \right) \;,
\end{equation}
and the normalization constant is set by $\int\de z\,\phi_i(z) = 1$.
$z_i^-$, $z_i^+$ are the photometric redshifts defining the edges of
the $i$th bin. We consider the following redshift bins, both with
surface density of galaxies $\bar n_g = 3\ {\rm arcmin}^{-2}$
(shot-noise $\epsilon = 1/\bar{n}_g \approx 2.8 \times 10^{-8}
\textrm{ sr}$):
\begin{itemize}
\item Low redshift $[0.42, 0.56]$.
\item High redshift $[0.90, 1.02]$.
\end{itemize}

We assume $b_1=1.5$ and set the non-linear coefficients
$b_2\approx-0.69$ and $b_{s^2}\approx-0.14$ according to the
fitting formula (based on $\Lambda$CDM simulations)
\cite{Lazeyras:2015lgp}
\begin{equation}
  \label{eq:b2}
  b_2 = 0.412 - 2.143\ b_1 + 0.929\ {b_1}^2 + 0.008\ {b_1}^3 \;,
\end{equation}
valid in the range $1 \lesssim b_1 \lesssim 9$. We assume Lagrangian
local-in-matter-density bias model
\begin{equation}
  \label{eq:bs}
  b_{s^2}= -\frac{2}{7}(b_1-1) \;,
\end{equation}
reviewed in \cite{Desjacques:2016bnm}. The precise value of the bias
coefficients is not relevant for our purposes.

\subsection{Spectroscopic surveys}
We consider a low-redshift SKA1-like neutral hydrogen galaxy survey
and a high-redshift \euc-like spectroscopic survey. Given the good
spectroscopic redshift determination, $W_i(z)$ is well approximated
by a tophat within the given redshift bins. Also here we consider a
low redshift and a high redshift survey, chosen to compare roughly
with the photometric survey bins:
\begin{itemize}
\item SKA1: $z \in [0.4, 0.6]$, with shot-noise
  $\epsilon=1/\bar{n}_g\approx 1.45 \times 10^{-5}$ sr and linear
  galaxy bias $b_1\approx 1.02$ consistent with SKA1 Medium-Deep Band
  2 Survey ($5\sigma$ detection threshold)
  \cite{Bacon:2018dui,Tanidis:2019teo}. We use again \autoref{eq:b2}
  and \autoref{eq:bs} as galaxy bias prescription. Given the smaller
  redshift range covered than the photometric case, here we neglect
  the redshift evolution of $\de N/\de z/\de\Omega$ when integrating
  over selection functions.
\item \textit{Euclid}: $z \in [0.9, 1.1]$, with shot-noise
  $\epsilon=1/\bar{n}_g\approx 1.68 \times 10^{-7}$ sr consistently with
  \cite{Blanchard:2019oqi}, and galaxy bias
  \cite{Yankelevich:2018uaz,DiDio:2018unb}
  \begin{align}
    \label{eq:7}
    b_1(z) &= 0.9 + 0.4 z\\
    b_2(z) &= -0.704172 - 0.207993 z + 0.183023 z^2 - 0.00771288 z^3 \;,
  \end{align}
  computed at the redshift bin mean $\bar{z}$. Again, $b_{s^2}$ is given
  by \autoref{eq:bs}. Following \cite{Blanchard:2019oqi}, also in
  this case we assume a constant $\de N/\de z/\de\Omega$ when
  integrating over selection functions.
\end{itemize}
Redshift bins have been chosen so far to match typical Fourier-space
galaxy clustering configurations, which are not expected to be optimal
for harmonic-space studies. While a fully tomographic analysis is
outside the scope of the present work, we also consider a narrower
high-redshift bin:
\begin{itemize}
\item \textit{Euclid}: $z \in [0.99, 1.01]$, with shot-noise
  $\epsilon=1/\bar{n}_g\approx 1.68 \times 10^{-6}$ sr. All other
  specifications are taken to be the same as in the last item
  above.
\end{itemize}

\section{Forecast methodology and results}
\label{sec:forecast}

\iffalse
(require 'cosmo)

(defun get-lmax (z)
  "Compute lmax at given Z."
  (let* ((h (/ (cosmo-get-hubble 0) 100.0))
         ;; The original 0.2 factor from arXiv:1907.02975 is too
         ;; optimistic for the bispectrum, divide it by 2.
         (kmax (* 0.1 h (expt (+ 1 z) (/  2.0 3.0)))))
    (* kmax (cosmo-get-los-comoving-distance z))))

(defun get-rmin (lmax z)
  "Compute smallest scale at given LMAX and Z."
  (* (cosmo-get-los-comoving-distance z) (/ (* 2 float-pi) lmax)))

(get-lmax 1.0)
(get-lmax 0.5)

(get-rmin 370 0.9)
(get-rmin 200 0.4)
\fi

In this section we forecast bispectrum detection perspectives. Due to
our tree-level bispectrum approximation, we only consider mildly
non-linear scales in the following analysis. For the lower redshift
bins (mean redshifts $\bar z \sim 0.5$) we set $\ell_{\rm max}$ values
up $\ell_{\rm max} = 200$, corresponding to transverse scales of about
$r(z=0.4) \approx 50$ Mpc at our lowest, most non-linear, redshift bin
edge.\footnote{We use $r(z) \approx d(z)\theta(\ell)$, where $d(z)$ is
  the line-of-sight comoving distance and $\theta(\ell) = 2\pi/\ell$
  \cite[e.g.][]{DiDio:2013sea}.} For the higher redshift bins ($\bar z
\sim 1$) we set $\ell_{\rm max}$ values up to $\ell_{\rm max} = 300$,
corresponding to $r(z=0.9) \approx 65$ Mpc at the lowest redshift
bin.\footnote{\label{fn:lmax} The largest wave number reachable with
  our perturbative treatment can be estimated as $k_{\rm max}(z) =
  0.1h(1+z)^{2/(2+n_s)}$ \cite{Maartens:2019yhx}, giving $\ell_{\rm
    max} \approx 170, 370$ at $z \approx 0.5, 1$, respectively. Our
  $\ell_{\rm max}=300$ value at $z \approx 1$ is also set by
  computational requirements for the case where we evaluate all of the
  multipoles to test our methodology, as in principle smaller scales
  could be reached compared to lower redshifts. It should also be
  reminded that, due to the different redshifts involved when
  integrating over selection functions, the correspondence of a given
  triangle in multipole and configuration space is not trivial. In
  actual observational analyses, the maximum multipole $\ell_{\rm
    max}$ should be set based on agreement with simulations tailored
  to the particular survey.}

We estimate the cumulative SNR up to $\ell_{\rm max}$, given in
\autoref{eq:csn}. The required number of multipole configurations
satisfying \autoref{eq:triangle-1} and \autoref{eq:triangle-2} are
$347,755$ for $\ell_{\rm max}=200$, and $1,157,880$ for $\ell_{\rm
  max}=300$. This is computationally prohibitive when including
observational selection functions (see \autoref{sec:numeric}).
Contrary to the case of the power spectrum, where 1-dimensional spline
interpolation over one multipole $\ell$ is routinely used to achieve
speedups of a factor 5--10 inducing errors well below $\lesssim 1\%$
\cite{Blas:2011rf}, 3-dimensional interpolation over the $(\ell,
\ell^\prime, \ell^{\prime\prime})$ multipole triplet is no longer
efficient enough. Instead, we approximate
\begin{equation}
  \label{eq:sn_mean}
  \SNR(\le \ell_{\rm max})
  \approx
  \sqrt{n_{\rm tot} \frac{1}{n_{\rm p}}
    \sum_{j=1}^{n_{\rm p}} \left(\frac{B_{j}^2}{\sigma^2_{B_{j}}}\right)}
  \;,
\end{equation}
where $j$ denotes a given $(\ell, \ell', \ell'')$ multipole
configuration. In other words, we approximate the arithmetic mean over
all $n_{\rm tot}$ physical configurations with the one over a partial
subset of $n_{\rm p}$ configurations randomly drawn from the total
ones. This allows us to recover $\SNR(\le\ell_{\rm max})$ at the
$\mathcal{O}(1\%)$ level considering only a few ($n_{\rm p} \sim
10^3$) configurations. We refer the reader to \autoref{sec:interp},
where we compare SNR estimates obtained with a different methodology.

\begin{figure}[t]
  \includegraphics[scale=0.45]{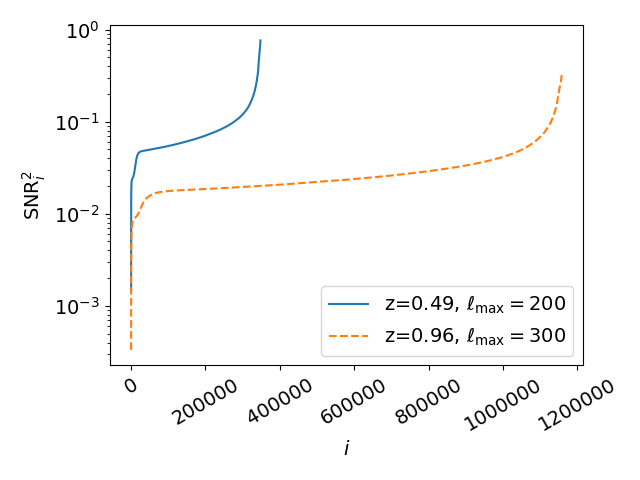}
  \includegraphics[scale=0.45]{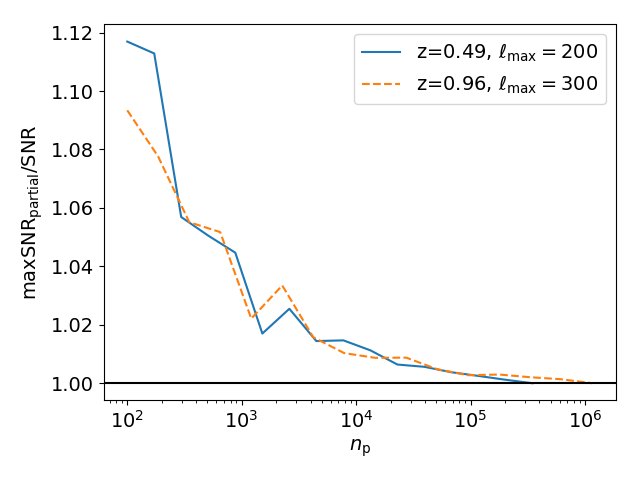}
  \caption{\emph{Left panel:} Sorted SNR for the cases where we
    neglect radial selection functions. Each index $i$ corresponds to
    a multipole configuration $(\ell, \ell', \ell'')$. \emph{Right
      panel:} We estimate the error induced by approximating the
    cumulative SNR considering only a partial subset of $n_{\rm p}$
    multipole configurations. For each $n_{\rm p}$ we consider 100 different
    random draws from the full set of $n_{\rm tot}$ multipole
    configurations, and we plot the largest deviation compared to the
    exact result.}
  \label{fig:sn_mean_err}
\end{figure}

To validate the methodology, we first consider the cases without
radial selection functions for which we can compute the cumulative SNR
using all the multipole configurations. In \autoref{fig:sn_mean_err}
we arrange triangle configurations $(\ell, \ell', \ell'')$ to show the
sorted SNR. The plot suggests that most configurations have comparable
SNR, 1--2 order of magnitudes smaller than the larger SNR. Hence, the
cumulative SNR cannot be well approximated considering only the
largest SNR configurations (folded multipole triangles, see
\autoref{sec:geom}). However, this also suggests that using only a
subsample of triangles to estimate the cumulative SNR is not sensitive
to missing large-SNR configurations. In the right panel of
\autoref{fig:sn_mean_err} we approximate the cumulative SNR as in
\autoref{eq:sn_mean}. We compute deviations with respect to the
non-approximate cumulative SNR considering 100 different random
selections of the partial subset of $n_{\rm p}$ triangles for each
$n_{\rm p}$ and show the largest deviation for each $n_{\rm p}$. This
gives an estimate of systematic errors introduced by our methodology,
mitigating the risk of underestimating them due to a particular random
draw. We expect to recover the cumulative SNR within $\sim 10\%$ for
$n_{\rm p} \gtrsim 200$, within $\sim 5\%$ for $n_{\rm p} \gtrsim
10^3$, and within $\sim 1\%$ for $n_{\rm p} \gtrsim 10^4$.

\begin{figure}[t]
  \includegraphics[scale=0.45]{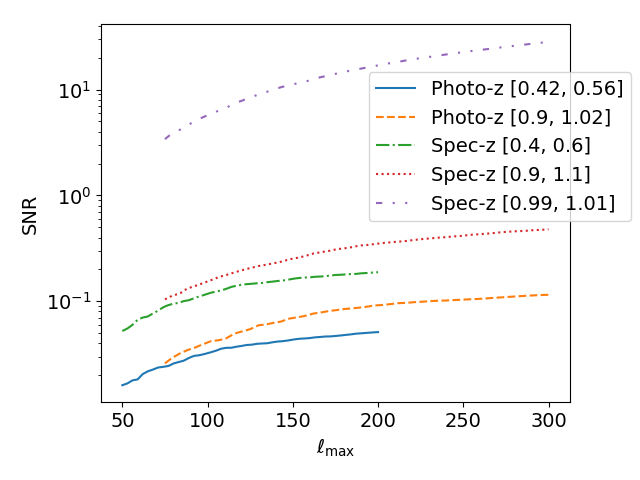}
  \includegraphics[scale=0.45]{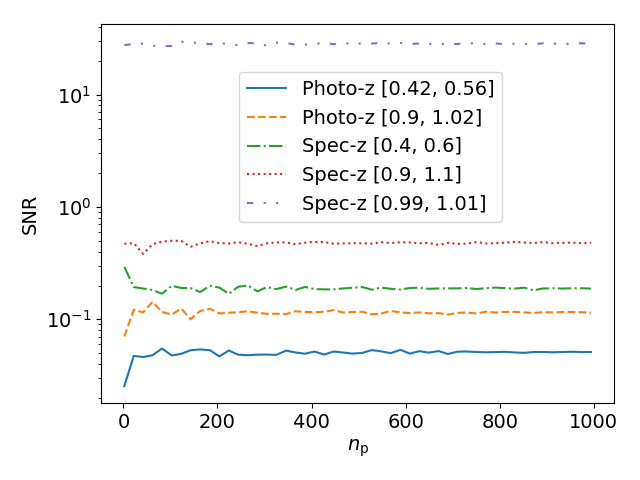}
  \caption{\emph{Left panel:} Cumulative SNR as a function of the
    maximum multipole for our reference surveys. \emph{Right panel:}
    Convergence test showing the cumulative SNR as a function of the
    number of points used to estimate the mean in
    \autoref{eq:sn_mean}.}
  \label{fig:csn_mean}
\end{figure}

The cumulative SNR as a function of $\ell_{\rm max}$ is shown in
\autoref{fig:csn_mean} for our reference surveys. In each case we
consider at least $n_{\rm p} = 1000$ multipole configurations to
estimate \autoref{eq:sn_mean}. The curves show small SNR values at
small $\ell_{\rm max}$ due to cosmic variance and grow roughly
linearly (note the logarithmic scale in the figure).\footnote{For
  comparison, neglecting shot-noise the linear power spectrum
  cumulative SNR also grows linearly $\sqrt{\sum_{\ell=2}^{\ell_{\rm
        max}} \left( \ell + 1 / 2 \right)} \approx \sqrt{1/2}\ell_{\rm
    max}$ for $\ell_{\rm max} \gg 1$
  \cite[e.g.][]{DiDio:2013sea,Tanidis:2019teo}.} Excluding for the
moment the spectroscopic case at $\bar z \sim 1$ with narrower width
$\Delta z = 0.02$, the largest SNR are in the range 0.05--0.5.
Spectroscopic bins outperform photometric ones due to the fact that
large photometric bins significantly smooth and reduce the signal
compared to the relatively narrow spectroscopic bins. Furthermore,
high mean redshifts $\bar z \sim 1.0$ bring larger SNR by about a
factor 2 than lower $\bar z \sim 0.5$ (despite the fact that we
include smaller comoving scales at $\bar z \sim 0.5$ due to our
choices of $\ell_{\rm max}$ dictated by computational limit, as
commented in footnote~\ref{fn:lmax}). On the one hand, gravitational
non-linearities lead to a larger bispectrum at low redshifts
\cite[e.g.][]{DiDio:2018unb}. On the other hand, in the spectroscopic
case the $\bar z \sim 0.5$ bin is affected by a shot-noise 2 orders of
magnitude larger than the $\bar z \sim 1.0$ bin, and the same bin
width $\Delta z=0.2$ corresponds to smoothing the signal over larger
comoving scales at $\bar z \sim 0.5$. In the photometric case both
bins have the same shot-noise, but due to the $\de N/\de z/\de\Omega$
distribution this comes at the cost of a significantly larger bin at
$\bar z \sim 0.5$ than at $\bar z \sim 1.0$.

Let us now focus on the $\bar z \sim 1$ case with bin width $\Delta z
= 0.02$. Although it is characterized by a larger shot-noise than the
spectroscopic $\Delta z = 0.2$ bin at $\bar z \sim 1$, the balance
with the largest signal (a consequence of the narrower bin) increases
drastically the SNR up to $\sim 30$.\footnote{We verified that SNRs
  eventually decrease considering even narrower window functions due
  to the increase of shot-noise relevance. For instance, the SNR for a
  $\Delta z = 10^{-4}$ bin width is about a factor of 2 smaller than
  our $\Delta z = 0.02$ case. Note that this comparison is only meant
  as a consistency check: spectroscopic redshift errors $0.001(1+z)$
  prevent analyses within $\Delta z = 10^{-4}$ bins.} This confirms
that typical configurations used for Fourier-space galaxy clustering
analyses should be revised for analyses in harmonic-space. It also
proves that the bispectrum will be detectable with upcoming galaxy
surveys.

While the bispectrum is expected to be more degraded by discretness
effects than the power spectrum \cite{Chan:2016ehg}, shot-noise does
not dominate the signal for our cases. It contributes significantly
($\sim 20\%$ of the variance) only in the spectroscopic case at $\bar
z \sim 0.5$ and it is subdominant ($\lesssim 1\%$ of the variance) for
other configurations. Also note that our choice for $\ell_{\rm max}$
is very conservative for the cases at $\bar z \sim 1$. However, even
assuming that shot-noise remains subdominant, extrapolating the
roughly linear growth of \autoref{eq:sn_mean} in
\autoref{fig:csn_mean} up to a more realistic $\ell_{\rm max} \sim
370$ (see footnote \ref{fn:lmax}) will not affect qualitatively our
conclusions. Depending on shot-noise balance, non-linearities beyond
our tree-level treatment will further boost the signal. Also note that
while neglecting redshift-space distortions as outlined in
\autoref{sec:bisp} is accurate for $\Delta z \sim \mathcal{O}(0.1)$
redshift bins, their inclusion is expected to enhance the cumulative
signal-to-noise for the $\Delta z = 0.02$ bin by $\mathcal{O}(10\%)$
\cite{Durrer:2020orn}.

As a convergence test, in \autoref{fig:csn_mean} we also show the
cumulative SNR for the largest $\ell_{\rm max}$ value in each case, as
a function of the number $n_{\rm p}$ of points used to estimate the
mean in \autoref{eq:sn_mean}. Results converge within $\sim 1 \%$
towards the largest $n_{\rm p}$, consistently with the error analysis
in \autoref{fig:sn_mean_err}.

\section{Data compression}
\label{sec:kl}

Here we discuss how to achieve a drastic dimensionality reduction in
data fitting analyses using the KLT. Let $\bm x$ be a Gaussian
distributed $n$-dimensional data vector, and let $\langle{\bm
  x}\rangle$ depend on the $m$-dimensional parameters vector
$\bm\theta$ that we want to constrain. The likelihood and Fisher
matrix are defined by
\begin{align}
  \label{eq:5}
  \log \mathcal{L} &\propto \frac{1}{2} ({\bm x} -
  \langle{\bm x}\rangle)^\textrm{\textsf{T}} {\bm C}^{-1} ({\bm x} -
  \langle{\bm x}\rangle)\; ,\\
  F_{ij} &= \frac{1}{2} \Tr\left[ {\bm C}^{-1}{\bm
     C}_{,i}{\bm C}^{-1}{\bm C}_{,j} + {\bm C}^{-1}
     \left(\langle{\bm x}\rangle_{,i} \langle{\bm
     x}\rangle_{,j}^{\ t} + \langle{\bm x}\rangle_{,j}
     \langle{\bm x}\rangle_{,i}^{\ t}\right)\right] \;,
\end{align}
The covariance and derivatives entering the Fisher matrix are
evaluated at a fiducial cosmology. The KLT is a linear transformation
that compresses, without information loss (in the Fisher matrix), the
$n$-dimensional data vector into a $m$-dimensional one. Then,
parameters can be constrained based on a likelihood that depends on
the compressed data set, and on a $m \times m$ covariance matrix
(rather than the original $n \times n$ one). This dimensionality
reduction in the covariance matrix is the main advantage of the KLT
(there is no advantage for the computation of the theoretical model).
In the case of the bispectrum we expect a large improvement given $m
\ll n$.

Let ${\bm A}$ be a $m \times n$ transformation matrix, and
${\bm y}$ the $m$-dimensional compressed data vector, i.e.\
\begin{equation}
  \label{eq:1}
  {\bm y} = {\bm A}\, {\bm x}
\end{equation}
Suppose we are only interested in one parameter, $m=1$.\footnote{To
  analyse joint constraints on $m>1$ parameters one can follow the
  MOPED algorithm \cite{Heavens:1999am}, or diagonalise the Fisher
  matrix (e.g.\ via PCA) before compressing \cite{Gualdi:2017iey}.}
Let ${\bm a}^\textrm{\textsf{T}}$ be the only non-vanishing row of the
${\bm A}$ matrix. Then, the Fisher matrix has one entry that we
label $i$
\begin{equation}
  \label{eq:2}
  F_{ii} = \frac{1}{2} \left( \frac{{\bm a}^\textrm{\textsf{T}}
      {\bm C}_{,i} {\bm a}}{{\bm a}^\textrm{\textsf{T}} {\bm
        C} {\bm a}} \right)^2 + \frac{\left({\bm a}^\textrm{\textsf{T}}
      \langle{\bm x}\rangle_{,i} \right)^2}{{\bm a}^\textrm{\textsf{T}}
    {\bm C} {\bm a}} \;.
\end{equation}
We assume the covariance to be weekly dependent on the parameters,
such that the first term is negligible compared to the second one.
This is an approximation that works well in practical applications
\cite{Heavens:1999am, Gualdi:2017iey, Heavens:2017efz, Carron:2012pw,
  Kodwani:2018uaf}. Then, it can be shown analytically
\cite{Tegmark:1996bz} that $F_{ii}$ is maximised by
\begin{equation}
  \label{eq:3}
  {\bm a}_i = {\bm C}^{-1}\langle{\bm
    x}\rangle_{,i} \;,
\end{equation}
which gives
\begin{equation}
  \label{eq:4}
  y_i = {\bm a}^\textrm{\textsf{T}} \langle{\bm x}\rangle_{,i} =
  \langle{\bm x}\rangle_{,i}^{\ t} {\bm C}^{-1}
  \langle{\bm x}\rangle_{,i} \;.
\end{equation}
Derivatives are taken at a fiducial cosmology.

Inference can be carried out considering the likelihood or Fisher
matrix of the compressed data
\begin{align}
  \label{eq:KL_C}
  \log \mathcal{L} &\propto \frac{1}{2} ({\bm y} -
  \bar{\bm y})^\textrm{\textsf{T}} \left[{\bm
     a}_i^{\ t} {\bm C} {\bm a}_j\right]^{-1} ({\bm y} -
  \bar{\bm y})
  \;,
  \\
  \label{eq:KL_F}
 F_{ii} &= \langle{\bm x}\rangle_{,i}^{\ t} {\bm
     C}^{-1} \langle{\bm x}\rangle_{,i} \;.
\end{align}

The formalism can readily be applied to the bispectrum detection. Our
$n$-dimensional data vector and covariance respectively read
\begin{align}
{\bm x} &= \{B_{\ell_1\ell^\prime_1\ell^{\prime\prime}_1},\,
B_{\ell_2\ell^\prime_2\ell^{\prime\prime}_2},\, \ldots,\, B_{\ell_n\ell^\prime_n\ell^{\prime\prime}_n}\}\;,\\
{\bm C} &= {\rm diag}\left(\sigma_{\ell_1\ell^\prime_1\ell^{\prime\prime}_1}^2,\,
  \sigma_{\ell_2\ell^\prime_2\ell^{\prime\prime}_2}^2,\, \ldots,\,
  \sigma_{\ell_n\ell^\prime_n\ell^{\prime\prime}_n}^2 \right)\;.
\end{align}
A detection analysis can
be formalised in terms of constraining the overall amplitude $\theta$
of the data ${\bm x} = \theta \tilde{\bm x}$. Then
$\langle{\bm x}\rangle_{,\theta} = \theta^{-1}
\langle{\bm x}\rangle$ and our fiducial parameter is
$\theta=1$. At tree-level the bispectrum covariance is computed
assuming $\langle B_{\ell\ell^\prime\ell^{\prime\prime}} \rangle \approx 0$, so it is
independent of the amplitude of non-Gaussian coefficients,
${\bm C}_{,\theta} \approx 0$. This allows us to estimate the
compressed covariance and Fisher matrix as given by
\autoref{eq:KL_C} and \autoref{eq:KL_F}.

\section{Conclusions}
\label{sec:conclusions}

In this work we discussed detection prospects of the gravitational
harmonic space bispectrum for upcoming galaxy surveys. We consider
mildly non-linear scales where tree-level standard cosmological theory
is valid. First, to get insights about geometrical properties, we
studied the dependence of the gravitational bispectrum and its
variance on multipole triangles $\ell \leq \ell^\prime \leq
\ell^{\prime\prime}$ when neglecting observational radial selection
functions and setting equal redshifts $z=z'=z''$. We showed that the
SNR is peaked for folded triangles $\ell = \ell^\prime =
\ell^{\prime\prime}/2$, and minimum for equilateral triangles $\ell =
\ell^\prime = \ell^{\prime\prime}$.

The maximum multipole $\ell_{\rm max}=300$ included in the analysis
corresponds to $\mathcal{O}(10^6)$ physical multipole triangles. We
showed how to estimate the cumulative SNR including observational
effects, in particular computationally expensive radial selection
functions, based on a partial subset of $\sim 1000$ multipole
configurations. We consider the complementary scenarios of high
redshift accuracy, low number density spectroscopic observations and
lower redshift accuracy, high number density photometric measurements
for cosmological galaxy surveys. As working assumptions, we adopt
\euc-like (both spectroscopy and imaging) survey and SKA1-like (line
galaxy) survey specifications. Specifically, we study redshift bins
with mean redshifts $\bar z \sim 0.5$ and $\bar z \sim 1$ for a
\euc-like photometric survey, compared to a spectroscopic bin at $\bar
z \sim 0.5$ for a SKA1-like survey, and a bin at $\bar z \sim 1$ for a
\euc-like spectroscopic survey. Considering redshift bin widths
$\Delta z = \mathcal{O}(0.1)$ consistent with typical galaxy
clustering configurations we show that, for a given redshift bin, the
spectroscopic measurements outperform the photometric ones.
Furthermore, bins at $\bar z \sim 1$ outperform those at $\bar z \sim
0.5$ by about a factor 2. For the spectroscopic surveys this is due to
a factor $\sim 100$ of difference in shot-noise. For the adopted \euc\
photometric survey specifications, all bins have the same shot-noise,
but given the galaxy selection function this implies a much wider
redshift bin at $\bar z \sim 0.5$ that smooths out and reduces the
signal.

Cumulative SNR values range between $\sim 0.05$ for the photometric cases, and
up to $\sim 0.5$ for the spectroscopic ones with widths $\Delta z =
\mathcal{O}(0.1)$. However, for comparison, a spectroscopic bin at mean
redshift $\bar z = 1$ and width $\Delta z = 0.02$ leads to a drastically
larger SNR $\sim 30$ thanks to optimal bin width and shot-noise balance,
suggesting that the bispectrum is detectable even for single bin analyses. We
have neglected partial sky coverage effects, but at first approximation the
cumulative SNR scales as $\SNR \to \sqrt{f_{\rm sky}}\ \SNR$
\cite{Durrer:2020orn}, where $f_{\rm sky}=0.3, 0.5$ are the sky fractions
covered by an \euc-like and SKA1-like survey, respectively. This decreases our
largest SNR to $\sim 15$, which is still very promising especially in view of
tomographic studies. For instance, the redshift resolution of an \euc-like
spectroscopic survey allows up $n_{\rm bin} \sim 100$ redshift bins in the
range $0.9 < z < 1.8$, which would translate into $n_{\rm bin} \times n_{\rm
  bin} \times n_{\rm bin}$ correlations; as computational costs soon increase
with the number of bins, methods to trim the number of \textit{cross-bin}
correlations should be considered \citep[see][]{Camera:2018jys}. The
tomographic analysis may also change conclusions about the relative
performance of photometric and spectroscopic surveys and given enough
tomographic bins harmonic-space statistics is expected to recover similar
information as Fourier-space statistics \cite{DiDio:2013sea,Nicola:2014bma}.
The inclusion of redshift cross-correlations may also change the SNR
dependence on triangles geometry as other effects here negligible, such as
lensing (our harmonic-space formulation makes it simple to include such
terms), will become relevant
\cite{DiDio:2014lka,DiDio:2015bua,Montanari:2015rga}. An even larger SNR will
be reached by including highly non-linear scales here neglected given our
tree-level approach, but our results show that mildly non-linear scales
already contain valuable information. This is confirmed in a parallel work
relying on the forecasting methodology here proposed \cite{Durrer:2020orn}
showing that the bispectrum of 21cm intensity maps (allowing very fine
redshift determination without being shot-noise limited, but whose
instrumental noise hinders small scales) will also be detectable.

We do not use the Limber approximation because for the harmonic
bispectrum it is not accurate even at relatively large multipoles
\cite{DiDio:2018unb}. This is computationally requiring given our
approach of estimating bispectra via integrations along the
line-of-sight. However, the expressions considered here are fully
compatible with more computationally efficient power-law expansions
\cite{Assassi:2017lea,Lee:2020ebj} that should be considered for future
development in this direction. It has been shown that replacing
line-of-sight integrals with such an expansion improves runtime up to
a factor 400 for the harmonic power spectrum
\cite{Schoneberg:2018fis}, hence presumably even more for the
bispectrum.

From a data fitting perspective, binning in multipole space and
efficient bispectrum estimators \cite[e.g.][]{Coulton:2017crj} should
be considered to afford bispectrum measurements given the large number
$\mathcal{O}(10^5)-\mathcal{O}(10^6)$ of multipole configurations here
considered. Simulations are needed to validate the non-linear scale
cutoff and the effects of finite survey volume neglected here. Given
that estimating covariance matrices from simulations would be
computationally prohibitive, we discussed how the Karhunen-Lo\`eve
transform let us to compress our $n$-dimensional data vector into a
single parameter, requiring the estimate of a $1 \times 1$ covariance.
The procedure can be extended to infer multiple parameters
\cite{Heavens:1999am}, and to compress as well radial modes in
tomographic analyses \cite{Alonso:2017hhj}. We leave a detailed
detection analysis applying our data compression framework to
simulations as a future development.

In this work we focused on the bispectrum induced by gravitational
evolution, useful to provide complementary constraints on standard
cosmological parameters \cite[e.g.][]{Gil-Marin:2016wya}. However,
the bispectrum is foremost a unique probe of primordial
non-Gaussianity. The primordial bispectrum can be comparable
to the gravitational one and both of them must be modeled jointly to
avoid systematic biases in parameter inference \cite{DiDio:2016gpd}.
Inclusion of the primordial bispectrum and a detectability analysis of
non-Gaussianity is then an important next step. The forecast methodology
outlined here can be applied to the total bispectrum induced by both
gravitational non-linearities and non-Gaussianity.

\acknowledgments

We thank Ruth Durrer, Mona Jalilvand, Rahul Kothari and Roy Maartens
for useful discussions. We acknowledge use of the Kerbero cluster at
IFT-UAM/CSIC (Madrid, Spain), and the Competence Centre for Scientific
Computing (C3S) and use of the OCCAM SuperComputer at Universit\`a
degli Studi di Torino (Turin, Italy). FM is supported by the Research
Project FPA2015-68048-C3-3-P [MINECO-FEDER] and the Centro de
Excelencia Severo Ochoa Program SEV-2016-0597. SC acknowledges support
from the Italian Ministry of Education, University and Research
(\textsc{miur}) through the `Departments of Excellence 2018-2022'
Grant (L.\ 232/2016) awarded by \textsc{miur} and Rita Levi Montalcini
project `\textsc{prometheus} -- Probing and Relating Observables with
Multi-wavelength Experiments To Help Enlightening the Universe's
Structure', in the early stages of this project.

\appendix

\section{Geometrical factors}
\label{sec:geomfact}
In this section we define geometrical quantities entering the
computation of the tree-level bispectrum, see \autoref{sec:bisp}.

The Gaunt integral is defined by
\begin{align}
  \label{eq:13}
  \mathcal{G}^{m_1m_2m_3}_{\ell_1\ell_2\ell_3}
  &=
      \int\de \Omega_{\bn}\, Y_{\ell_1m_1}(\bn) Y_{\ell_2m_2}(\bn) Y_{\ell_3m_3}(\bn)
  \\ \nonumber
  &= \tj{\ell_1}{\ell_2}{\ell_3}{0}{0}{0}
      \tj{\ell_1}{\ell_2}{\ell_3}{m_1}{m_2}{m_3}
      \sqrt{\frac{(2\ell_1+1) (2\ell_2+1) (2\ell_3+1)}{4\pi}}
      \;,
\end{align}
where $\Omega_{\bn}$ is the solid angle spanned by $\bn$ and we
introduced Wigner's 3-j symbols. The Gaunt integral satifies the
symmetries discussed in \autoref{eq:triangle-1} and
\autoref{eq:triangle-2} and the paragraph above them. The factor
\begin{equation}
  \label{eq:12}
  g_{\ell_1\ell_2\ell_3}=\sqrt{\frac{(2\ell_1+1)(2\ell_2+1)(2\ell_3+1)}{4\pi}}
  \tj{\ell_1}{\ell_2}{\ell_3}{0}{0}{0}
\end{equation}
relates the reduced bispectrum to the angle-averaged one
\begin{align}
  \label{eq:aa-bisp}
  g_{\ell_1\ell_2\ell_3} b_{\ell_1\ell_2\ell_3}(z_1, z_2, z_3)
  &=
      \sum_{m_1, m_2, m_3} \tj{\ell_1}{\ell_2}{\ell_3}{m_1}{m_2}{m_3}
      B_{\ell_1\ell_2\ell_3}^{m_1m_2m_3}(z_1, z_2, z_3)
  \\ \nonumber
  &=
      B_{\ell_1\ell_2\ell_3}(z_1, z_2, z_3)\;.
\end{align}
The factor
\begin{align}
\label{eq:Q_def}
Q_{\ell\ell^{\prime }\ell^{\prime \prime }}^{\ell_1\ell_2\ell_3}
=
  I_{\ell\ell^{\prime }\ell^{\prime \prime }}^{\ell_1\ell_2\ell_3}
  \Gj{\ell_1}{\ell_2}{\ell_3}{\ell^{\prime}}{\ell^{\prime\prime}}{\ell}
  \left( -1\right) ^{\ell+\ell^{\prime }+\ell^{\prime \prime }} \;,
\end{align}
is expressed in terms of Wigner's 6-j symbols and of
\begin{equation}
  \label{eq:15}
  I_{\ell\ell^{\prime }\ell^{\prime \prime }}^{\ell_1\ell_2\ell_3} \equiv
  \sqrt{(4\pi
    )^3(2\ell_1+1)(2\ell_2+1)(2\ell_3+1)}
  \tj{\ell}{\ell^{\prime \prime}}{\ell_1}{0}{0}{0}
  \tj{\ell^{\prime}}{\ell}{\ell_2}{0}{0}{0}
  \tj{\ell^{\prime \prime}}{\ell^{\prime}}{\ell_3}{0}{0}{0}
  \;.
\end{equation}
Typically, using Wigner's symbols symmetries, only a few coefficients
of $Q_{\ell\ell^{\prime }\ell^{\prime \prime }}^{\ell_1\ell_2\ell_3}$
are non-vanishing for a given $\ell$.

\section{SNR estimate based on interpolation}
\label{sec:interp}

Here we discuss an alternative method to \autoref{eq:sn_mean} to
approximate the cumulative SNR using only a partial subset of $n_{\rm
  p}$ multipole configurations. We use the fact that the
$\SNR^2_{\ell\ell^\prime\ell^{\prime\prime}}$ can be monotonically
sorted as discussed in \autoref{sec:geom}: we map triplets $(\ell,
\ell^\prime, \ell^{\prime\prime})$ to an index $i$ whose order sorts
$\SNR_{\ell\ell^\prime\ell^{\prime\prime}}$ (see
\autoref{fig:sn_mean_err}). Then, we compute
$\SNR^2_{\ell\ell^\prime\ell^{\prime\prime}}$ for $n_{\rm p}$ randomly
selected $(\ell, \ell^\prime, \ell^{\prime\prime})$ triangle
configurations. To sum over all triangles contributing to the
cumulative SNR, we distribute uniformly the selected configurations
over the whole index $i$ range and interpolate. More precisely, we
draw $n_{\rm p}-2$ random integers from a uniform distribution within
the open interval $(1, n_{\rm tot})$, where $n_{\rm tot}$ is the total
number of physical multipole triangles corresponding to $\ell_{\rm
  max}$, and include the boundaries $i=1, n_t$.

\begin{figure}[t]
  \includegraphics[scale=0.45]{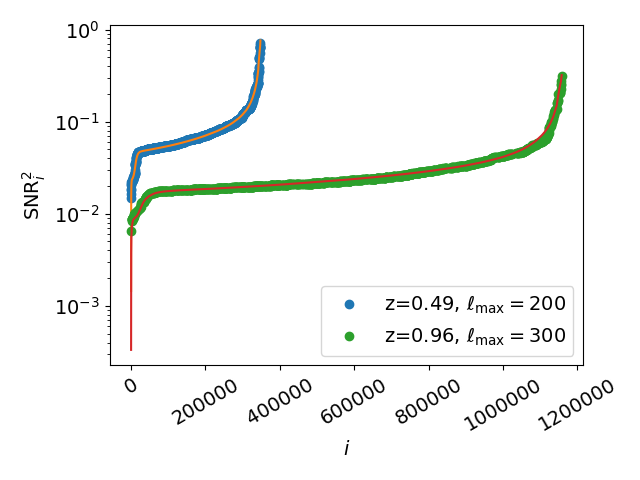}
  \includegraphics[scale=0.45]{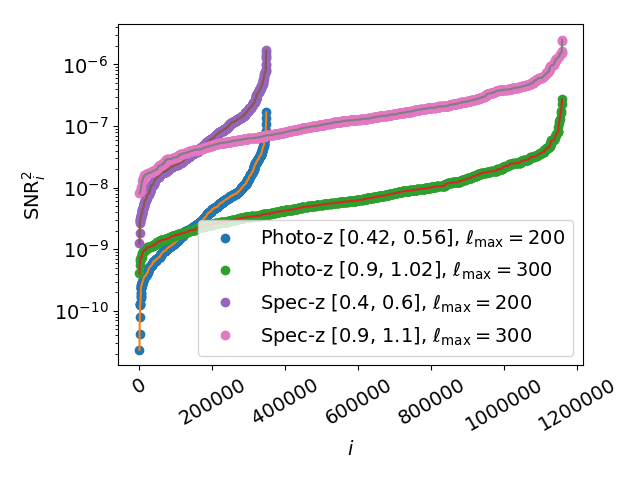}
  \caption{\emph{Left panel:} SNR for a random selection of $10^3$
    multipole configurations uniformely distributed along the whole
    range as described in the text. Solid lines show the full result
    for the cases that neglect radial selection functions. \emph{Right
      panel:} SNR per multipole configuration including our survey
    forecast specifications. Solid lines show the interpolating
    function.}
  \label{fig:sn_interp}
\end{figure}

This method is illustrated in \autoref{fig:sn_interp}. In the left
panel we compare the full result for the cases without selection
function to a random selection of $10^3$ interpolating configurations.
The tails of $\SNR_i$ are the most critical features driving sampling
requirements, together with the fact that $\SNR_i$ spans 2--3 order of
magnitude. We checked that linear and cubic interpolations agree well,
hence we opt for the simpler linear one. We verified that this method
agrees well with the one described in the main section, leading to
similar intrinsic systematic errors (see right panel of
\autoref{fig:sn_mean_err}).

The right panel of \autoref{fig:sn_interp} shows the $\SNR^2_i$
interpolation results for our reference surveys using the redshift
bins of width $\Delta z = \mathcal{O}(0.1)$. For each case we consider
at least $n_{\rm int} = 1000$. The functional dependence on the
sorting index $i$ is similar to case without selection function, hence
we expect to recover the cumulative SNR within 5\% errors. Results are
consistent with \autoref{fig:csn_mean}.

\section{Numerical computation}
\label{sec:numeric}

For the numerical computation of the bispectrum and its covariance we
use a modified version of the C++ backend of the Python-based
Byspectrum code originally developed in \cite{DiDio:2018unb}.
Cosmological transfer functions are computed using CLASS
\cite{Blas:2011rf}. We use the Suave algorithm of the Cuba library
\cite{Hahn:2004fe} to perform integrals over radial selection
functions,\footnote{We verified that trilinear interpolation of the
  integrand over the redshift grid $(z_1, z_2, z_3)$, see
  \autoref{eq:14}, is not efficient enough to bring significant
  improvements.} and the WIGXJPF library \cite{Johansson:2015cca} to
compute Wigner symbols required for the geometric terms defined in
\autoref{sec:geomfact}.

\begin{table}[t]
  \centering
  \begin{tabular}{|l|c|}
    \hline
    Configuration & Runtime \\
    \hline
    No $\phi(z)$ & 0.3 s \\
    Spectroscopic $\Delta z = \mathcal{O}(0.01)$ & 2 min\\
    Spectroscopic $\Delta z = \mathcal{O}(0.1)$ & 20 min\\
    Photometric $\Delta z = \mathcal{O}(0.1)$ & 60 min \\
    \hline
  \end{tabular}
  \caption{Average runtime required to compute the SNR for one
    multipole triangle configuration using 8 CPUs of an Intel Xeon CPU
    E5506 @ 2.13GHz processor. We include the cases without
    integration over selection functions $\phi(z)$, the spectroscopic
    and photometric redshift bins of our forecasts for different bin
    widths $\Delta z$.}
  \label{tab:runtime}
\end{table}

In table~\ref{tab:runtime} we report the average runtime to compute
$\SNR_{\ell\ell^\prime\ell^{\prime\prime}}$ for one multipole triangle
configuration for the different cases studied in this work, relative
to one node of a computer cluster.\footnote{Given the independence of
  the bispectrum at different multipole triangles, the computation can
  be further distributed over several nodes of the cluster.} This
table is only meant to provide an indicative order of magnitude. We
stress that runtime is not homogeneous across all triangle
configurations and, as described in \autoref{sec:forecast}, the cases
at larger mean redshifts $\bar z \sim 1$ reach larger multipole values
$\ell_{\rm max}=300$, compared to $\ell_{\rm max}=200$ at lower
redshifts $\bar z \sim 0.5$. The different $\ell_{\rm max}$ is the
main reason why, for each separate case reported in
table~\ref{tab:runtime}, computations at higher redshifts can take up
to 10\% longer than lower redshifts for comparable redshift bin
widths. Also, the scaling with the number of CPUs (within a single
node) is not linear, hence we report the runtime relative to all of
the CPUs used. In the cases without selection functions we compute the
bispectrum for different multipole triangles in parallel with
OpenMP.\footnote{\url{https://www.openmp.org/}} When including
selection functions, bispectra at different multipole triangles are
computed serially in a given node, but numerical integrals are carried
out in parallel as detailed above. In this case, the parallel
computation scheme \cite{Hahn:2014fua} does not lead to optimal CPU
loading. Furthermore, we consider a single set of numerical precision
parameters set to reach convergence for all cases, while each redshift
bin could be optimised separately leading to considerable speedup (as
hinted by the very strong dependence of runtime on the redshift bin
width). However, rather than improving on these aspects, we deem it
more promising to first pursue the power-law expansion mentioned in
\autoref{sec:conclusions} to significantly reduce runtime.

\bibliographystyle{JHEP}
\bibliography{biblio}

\end{document}